\documentclass[11pt]{article}
\usepackage{graphicx,hyperref}
\usepackage{amsmath, amssymb}
\usepackage{setspace}
\usepackage[dvips,letterpaper,margin=1.0in,bottom=1.0in]{geometry}
\pdfoutput=1
\title{\bf Gravitational Drainage of Thin Films of Trisiloxane-(Poly)ethoxylate Superspreaders}
\author{S. Sett, R. P. Sahu, S. Sinha-Ray and A.L. Yarin(ayarin@uic.edu)
\\
Department of Mechanical and Industrial Engineering,
\\
University of Illinois at Chicago,
\\
842 W. Taylor St., Chicago IL 60607-7022}
\date{}
\begin{document}
\begin{figure}[ht!]
\centering
\includegraphics[scale=0.4]{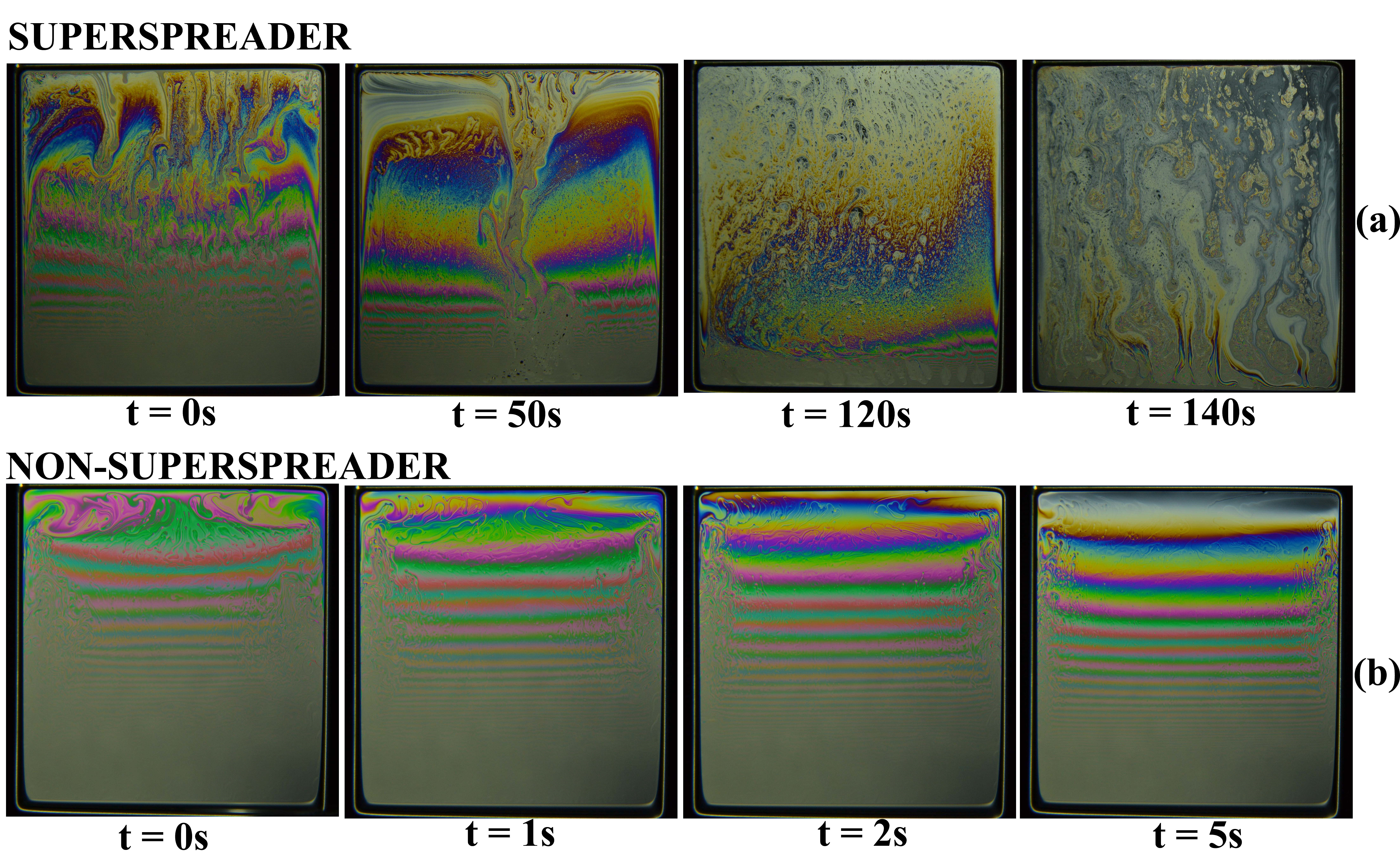}
\caption{Drainage of (a) superspreader Silwet L-77, and (b) its ``cousin''non-superspreader Silwet L-7607.}
\label{fig:combined-copy.png}
\end{figure}
\maketitle
The aqueous solutions of trisiloxane-(poly)ethoxylate surfactants have multiple applications as agricultural adjuvants, solid modifiers, and cleaners, as well as they are widely used in pharmaceutical and cosmetic industries. The trisiloxane surfactants are commonly denoted as M(D$'$E$_n$OMe)M, where M-D$'$-M represents the trisiloxane hydrophobe, E$_n$ is the ethylene oxide, Me stands for the methyl group, and O is oxygen~\cite{Venzmer-2011}. The superspreading ability of these surfactants when added to water drops on surfaces such as Teflon depends on the length of the (poly)ethylene oxide group. In respect to wetting hydrophobic Teflon surfaces, two drastically different ``cousin'' types of trisiloxane-(poly)ethoxylate surfactants with n=7.5 (which is the superspreader known under the product name Silwet L-77) and n=16 [which is a non-superspreader, denoted as M(\'D E$_n$)M under the product name Silwet L-7607] are recognized~\cite{Venzmer-2011}. However, their effect on gravitational drainage of vertical water films with no contact with any hydrophobic surface has never been studied~\cite{Sett-2013}, and is addressed here for the first time.

\begin{figure}[ht!]
\centering
\includegraphics[scale=0.35]{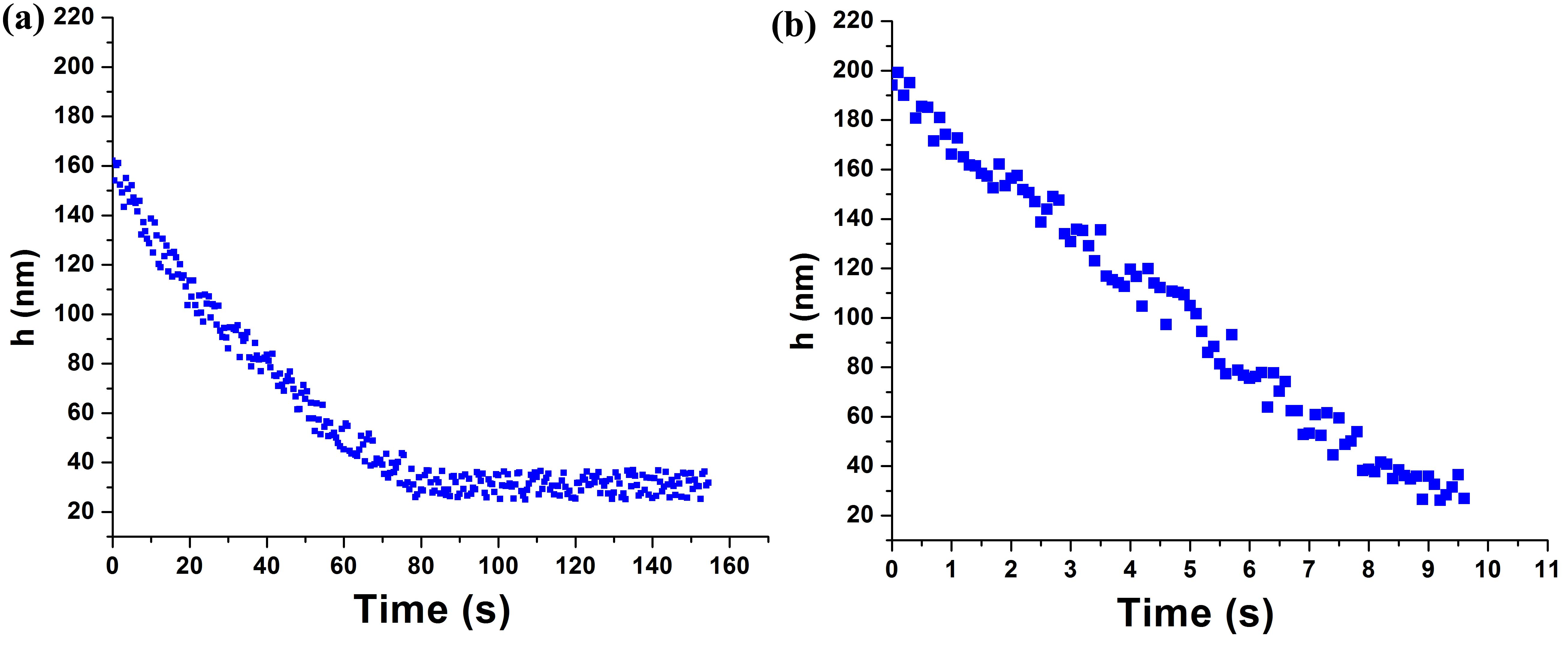}
\caption{Film thickness of (a) Silwet L-77 superspreader, and (b) its ``cousin'' non-superspreader Silwet L-7607. The data correspond to the film top.}
\label{fig:hvst-copy.png}
\end{figure}

Thin vertical films were formed from the 0.5 vol\% aqueous solutions of Silwet L-77 and Silwet L-7607. The drainage processes of the thin films formed by the superspreader (Silwet L-77) and the non-superspreader (Silwet L-7607) were radically different, both morphologically and by the duration~(Fig.\ref{fig:combined-copy.png}). The drainage dynamics of the thin films were observed using the microinterferometry technique~\cite{Sett-2013}. The thin films formed from the superspreader were stable for a much longer time in comparison with the ordinary surfactants~\cite{Sett-2013} and the non-superspreader, albeit also much more ``vigorous''~(Fig.\ref{fig:combined-copy.png}). The thinning of the superspreader film was practically arrested at the latter stage, whereas the ``cousin'' non-superspreader film continued to drain~(Fig.\ref{fig:hvst-copy.png}).

The thin film formed by the superspreader~(Fig.\ref{fig:combined-copy.png}a;{Video: Part1}) revealed a ``turbulent''-like motion visibly different from the marginal regeneration~(Fig.\ref{fig:combined-copy.png}). On the other hand, the non-superspreader film~(Fig.\ref{fig:combined-copy.png}b;{Video: Part2}) revealed the ordered drainage pattern characteristic of the ordinary surfactants~\cite{Sett-2013}.
In addition, the Dynamic Light Scattering (DLS) of the aqueous solution of the superspreader Silwet L-77 revealed the presence of the aggregates of sizes ranging from 1.6 nm to 800 nm, the latter being associated with the bilayer structures assumed for superspreaders in ~\cite{Venzmer-2011}. On the contrary, the DLS of the aqueous solution of the non-superspreader Silwet L-7607 revealed the presence of only small aggregates of the size about 3.5 nm, which are rationalized as ordinary micelles~\cite{Venzmer-2011}.
\\
The hydrophilic head group of the non-superspreader Silwet L-7607 is larger than that of the superspreader Silwet L-77, which thus favors the formation of spherical micelles in the former in distinction from the bilayer lamellae formed by the latter. The disjoining pressure associated with fluffy film surfaces resulting from the hanging bilayer lamellae is the physical source of the observed stabilization of the superspreader film drainage~(Fig.\ref{fig:hvst-copy.png}). In this case the disjoining pressure increases as p$_{disj} \sim 1/h^9$ (h is the film thickness) for 0.2-1 vol\% superspreader solutions, as h diminishes in the range below 100 nm.
\\
Similar results were obtained for another superspreader BREAK-THRU S 278, and its non-superspreader ``cousin'' BREAK-THRU S 233.
\\
We are grateful to the United States Gypsum for support of this work.

\end{document}